 \documentclass[preprint2]{aastex}

\slugcomment{Not to appear in Nonlearned J., 45.}

\shorttitle{The Effect of Strong Magnetic Field on the Standard
Model} \shortauthors{Jing-Jing et al.}

\begin{document}


\title{The Physical Essence of Pulsar Glitch}


\author{Qiu-He Peng \altaffilmark{1}, Jing-Jing Liu\altaffilmark{2} and Chih-Kang, Chou\altaffilmark{3}}
\email{qhpeng@nju.edu.cn, zsyuan@nao.cas.cn, and syjjliu68@126.com}


\altaffiltext{1}{Department of Astronomy, Nanjing University,
Nanjing, Jiangshu 210000, China} \altaffiltext{2}{College of Marine
Science and Technology, Hainan Tropical Ocean University, Sanya,
572022, China. Corresponding author E-mail:
syjjliu68@126.com}\altaffiltext{3}{National Astronomical
Observatory, Chinese Academy of Sciences, Beijing, 100000, China.}


\begin{abstract}
Based on the magnetic dipole radiation from the $^3P_2$ neutron
superfluid vortices ($^3P_2$ NSFV) in neutron stars, we propose a
model of glitch for young pulsars by oscillation between B phase and
A phase of $^3P_2$ Neutron superfluid. The main behavior of glitches
of pulsars may be naturally explained by our model. Our results show
that the glitch is a repeat phenomena with quasi period $^3P_2$
neutron superfluid B phase $\Longrightarrow $A phase
$\Longrightarrow $B phase $\Longrightarrow $ many repeated glitches
with quasi-period. With repeating of the phase transition, the
vortex quantum number, $n$, of $^3P_2$ NSFV is gradually reduced,
and the heating rate $\varepsilon ^{(B)}$ in B phase is also getting
lower and lower. After a number of glitches, the time intervals of
successive glitch will gradually become long, and the glitch
amplitude is downward. When the heating rate $\varepsilon ^{(B)}$ of
the old neutron star becomes lower than the cooling rate of the
Direct Urca, which can happen in super strong magnetic field
\citep{b45}, the B phase of the $^3P_2$ NSFV is no longer returned
to the A phase state. The phase oscillation of the system is stopped
immediately. That means, old pulsars will no longer present the
glitch. All of these results are consistent with observations
\citep{b34}. The slowing glitch phenomenon for some older pulsars is
a naturally result in our theory. The relationship between Glitch
amplitude and the stationary time interval (see Fig.2 from
\citet{b36}) is also naturally got by our theory.

\end{abstract}

\keywords{stars: neutron --- pulsars: general --- stars: evolution.}



\section {Introduction}
\subsection {Observations on glitches of pulsars}
The observed pulsar rotation periods often show that some young
pulsars experience one or more glitches (or macro jump). The regular
pulse signals could be occasionally shortened by glitches at typical
amplitude of $\Delta \Omega_0/\Omega_0\sim (10^{-10}- 10^{-6})$.
These glitches are usually accompanied by a spin-down effect at a
much larger rate $\Delta \dot{\Omega}/\dot{\Omega}\approx (10^{-3}-
10^{-2})$ \citep{b34,b7,b11}. The archetypal glitch neutron star is
the Vela pulsar, which has exhibited a regular sequence of similar
size glitches since the first observed event in 1969. There are 478
glitches detected among 174 pulsars up to date and 120 glitches
among them detected in eight of the glitch pulsars are great
glitches with $\Delta \Omega_0/\Omega_0> 10^{-6}$. Eleven glitches
with nine grand ones were detected from PSR Vela during 36 years .
There are also nineteen smaller glitches detected during 36 years
from PSR Crab with smaller magnitudes. In some pulsars, besides such
macro glitches, there are detected micro-glitches with jump
amplitudes less than $10^{12}$ in greater numbers.

The most important observational statistics of pulsar glitch
phenomena up to date are given as follows:

(1) There is a rough tendency for both the jump amplitude and the
frequency of glitches to decrease with the pulse period as the
pulsar ages \citep{b34}. Up to the present no glitches have been
detected in pulsars with periods longer than 0.7s.

(2) Pulsar glitch phenomena is primarily concentrated in the group
of very young pulsars with strong magnetic fields \citep{b34}: a)
The younger the pulsars the most often the glitch, the larger the
amplitude of the glitch becomes. The glitches are less often as the
pulsars age and the amplitude of the glitches are lowered. b) The
glitches are more often for pulsars with strong magnetic fields. On
the other hand, the glitches are less often for pulsars with weak
magnetic fields, and the lower the amplitudes of the glitches become
\citep{b35}. These observational facts can not be explained by
current pulsar models.

(3) Pulsar glitches are usually of served as a sudden change with
very short time scale. However, a slow glitch with long time scale
is more than several days in 2005. These observational facts can not
be explained by current pulsar models.

(4) \citet{b36} discovered that the young pulsar PSR J0573-6910
 in the Large Magellanic Cloud (LMC), which has glitch with amplitude
change roughly proportional to the time separation between two
successive glitches after ten years of observation to monitor the
radio pulsar period. This observation can not be explained by
current pulsar models either.

\subsection[]{Theoretical researches on the glitches of pulsars}

The consensus view is that these events are a manifestation of the
presence of a superfluid component in the star's interior
\citep{b49}. This idea was first put forward by \citet{b3}, who
envisaged a glitch as a tug-of-war between the tendency of the
neutron superfluid to match the spin down rate of the rest of the
star by expelling vortices and the impediment experienced by the
moving vortices due to pinning to crust nuclei. Strong vortex
pinning prevents the neutron superfluid from spinning down, creating
a spin lag with respect to the rest of the star (which is spun down
electro-magnetically). This situation cannot persist forever. The
increasing spin lag leads to a build up in the Magnus force exerted
on the vortices. Above some threshold pinning can no longer be
sustained, the vortices break free and the excess angular momentum
is transferred to the crust. This leads to the observed spin-up.

The current pulsar models for the mechanism to explain glitches are
given as follows:

(1) The star quake model \citep{b5}

This model predicts that the time separation between two successive
glitches for the pulsar Vela PSR is roughly 1000 years. This is
quite different from the observed facts of 36 years, eleven times.

(2) Vibration model of the neutron star core \citep{b46}

In this model, glitch may appear every several years with energy
release $10^{45}$erg. The neutron star is immediately heated to
become a strong X-ray source. However, such strong X-rays has never
been observed during or after glitch.

(3) A creep model of vortex filament by an action of shell -
superfluid coupling \citep{b1,b2,b3}

\textbf{Pulsar glitches are contributed to the motion of superfluid
vortex lines; this lines tend to be pinned to nuclei pf stellar
crust and sudden, large - scale creep of these lines from one
pinning site to another may be responsible for glitches
\citep{b241}. In this model, the key idea is that the roots of the
superfluid vortexes slide randomly in the inner shell, and
occasionally they are pinned to the heavy nucleus. This model has
been now regarded as the mainstream model by most researchers
concerned \citep{b19,b190,b191,b192,b193}. In their review paper
\citep{b190}, \citet{b190} wrote a following detail comment:
\citet{b3} suggested that interactions between vortices and ions in
the NS crust can 'pin' the vortices and restrict their outward
motion. As long as the vortices are pinned, i.e. stay fixed in
position, the superfluid does not spin down, storing angular
momentum, which is periodically released in glitches. The vortex
model has become the standard picture for pulsar glitches".
"Nevertheless several issues still need to be resolved before the
vortex picture attains the status of a self-consistent, predictive,
falsifiable theory. The trigger for vortex unpinning is still
unknown and may be due to vortex accumulation in strong pinning
areas, vortex domino effects, hydro-dynamical instabilities or
quakes. Recent calculations have also showed that Bragg scattering
severely limits the mobility of neutrons in the crust, limiting the
amount of angular momentum that can be stored in the crust between
glitches. An analysis of the Vela pulsar reveals that in this case
it is difficult to accommodate the implied angular momentum during a
glitch unless the pulsar has a low mass $\leq 1M_{\bigodot}$ or part
of the core is involved in the process "\citep{b190}.}

\textbf{This model may be the mechanism for the production of micro
glitches of neutron stars . But it is difficult for this model to
explain the grand glitches of Vela PSR. Besides, there are too many
free parameters in this model. It is rather difficult to determine
all these parameters to explain a series of observational facts of
neutron star glitches to be elaborated later. }

(4) A model due to the effect of twisted neutron superfluid vortex
filament with proton super-conducting flux tube \citep{b51}.

Although this model may explain the grand glitch ($\Delta
\Omega_0/\Omega_0\geq 10^{-6}$), but it contains too many free
parameters to determine, consequently, it is also difficult for this
model to explain the observed glitches. \citet{b11,b23,b48}
discovered, in view of the fact that the pulsar spin axis does not
coincide with its magnetic axis, the Two-component model (i.e.,
neutron superfluid vortices tangle with proton super-conducting
magnetic tulle model) would predict the procession of the neutron
star spin axis with periods of several seconds. Analysis of the
observed data for PSRB 1818-11 gives the procession period of the
neutron star spin axis of the order of several years. Thus, they
conclude that the neutron superfluid vortices may not coexist with
the type II proton super-conducting state (super-conducting magnetic
tube). \textbf{Although it might appear in the inner core of the
neutron star with density in the region of $2-3\rho_{\rm{nuc}}$
\citep{b191}, however, the quark model of nuclear matter in this
density region may be dominant. Therefore, \citet{b23} have series
doubts about this model.}

(5) \textbf{\citet{b361} investigated Kelvin-Helmholtz instability
and circulation transfer at an isotropic-anisotropic superfluid
interface in a neutron star. By hydrodynamic method, they suggested
that this instability may provider a trigger mechanism for pulsar
glitches. However, further explanations of many observed phenomena
of pulsar glitches are needed for their theory.}

Besides, \citet{b80,b81,b82} have conducted a series of researches
concerning the glitch phenomenal of the Vela pulsar. They considered
the coupling of the shell with  the superfluid interior of neutron
stars. They did not investigate the mechanism for the production of
glitches of pulsars.

Up to now, we note that the physical reason for the generation of
pulsar glitches is still not clearly understood and it is one of the
most difficult puzzling topic in pulsar researches. \textbf{In this
paper, we propose a new physical mechanism for generating Glitch,
which is completely different from the existing known models. The
main ideas of ours are mentioned in the abstract of the paper.}

This paper is organized as follows. In Section 2, we study the
anisotropic $^3P_2$ superfluid vortex motion in neutron star
interiors. In Section 3, we summarize our researches and discuss the
cooling and heating problem in neutron star interiors. In Section 4,
we present our model and analyze the properties of pulsar glitch.
Some discussions and conclusions are given in Section 5.


\section[]{The anisotropic $^3P_2$ superfluid vortex motion in neutron star interiors}

\subsection[]{Two types of neutron superfluid}

There are two types of superfluid with different properties between
the thin crust and the interior of neutron stars. For densities
$10^{11} < \rho(\rm{g/cm^3}) < 1.4\times10^{14}$, the superfluid is
isotropic. The energy gap of the Cooper pairs may reach (1-2) MeV in
the $^1S_0$ state. The initial surface temperature of the nascent
neutron star from the gravitational collapse of the supernovae core
may reach $10^{11}$K. In a relating short time scale the temperature
is lowered to roughly $10^{6}$K, for example, the Crab PSR was
formed in 1054 during supernovae explosion. The surface temperature
of the Crab has lowered to $10^{6}$K in less than a thousand year.
The interior temperature of it was estimated to be
$2.0\times10^{8}$K. When the interior temperature is lowered
critical temperature of the $^1S_0$ state, $T_c(^1S_0)=\Delta
(^1S_0)/k\approx 10^{10}$K (where k is the Boltzmann constant), the
isotropic neutron superfluid will appear. For neutron Cooper pairs
coupled by the $^1S_0$ wave interaction, the spins of the two
neutrons in the Cooper pair are anti-parallel, so that the total
spin of the $^1S_0$ Cooper pair is zero and there is no net magnetic
moment. The $^1S_0$ state is isotropic in external magnetic fields.
The property of this state is similar to liquid $^4$He ($^4$He II)
approaching near to absolute zero (lower than 0.2K) in the earth's
laboratories. When matter densities in the range $3.3\times 10^{14}<
\rho (\rm{g/cm^3})<5.2\times10^{14}$ (Note that nuclear density
$\rho_{\rm{nuc}}=2.8\times10^{14}\rm{g/cm^3}$), the energy gap of
the Cooper pairs in the $^3P_2$ state is roughly 0.045Mev
\citep{b10}. When the interior temperature of neutron state is
lowered to below $2.8\times10^8$K, the neutron fluid in this region
will become the anisotropic $^3P_2$ superfluid after a phase
transition from the normal fluid. For neutron Coupler Pairs coupled
by the $^3P_2$ wave interaction the spins are parallel, so that the
total spin is equal to 1 and there is equal a net magnetic is just
twice that of the neutron (abnormal magnetic moment).

In the presence of external magnetic field, the magnetic moment of
$^3P_2$ neutron Cooper has a tendency to reverse the direction of
the external magnetic field (temperature thermal effect makes the
magnetic moment of the $^3P_2$ neutron Cooper to be chaotic (i.e.,
in confusion) in direction due to thermal agitation). Consequently,
the $^3P_2$ state is anisotropic superfluid $^3P_2$ state. The
properties of this $^3P_2$ state is similar to the liquid $^3$He for
temperatures approaching absolute zero in the earth's laboratories
(lower than 0.02K).

\subsection[]{Neutron superfluid vortex motion in neutron stars}

Rotating Superfluid are quantized to become superfluid Vortex flow
(or eddy current or whirling fluid) \citep{b12} being analogous to
type II superconductivity, the neutron superfluid in the interior of
neutron stars is in a vortex state, i.e., there are plenty of vortex
lines (vortex filament). In general, the vortex filaments are
arranged in a systematic lattice, they are parallel to the axis of
rotation of the neutron star and as a whole they revolve around the
axis of rotation of the neutron star almost rigidly. The circulation
of each vortex filament intensity $\Gamma$ is quantized
\begin{equation}
 \Gamma=\oint \vec{V}\cdot d\vec{l}=n\Gamma_0,~~~~\Gamma_0=\frac{2\pi\hbar}{m_n}
 \label{1}
\end{equation}
where $n$ is a circulation quantum number of the vertex, $m_n$ is
the mass of a neutron, $\hbar$ is Planck's constant divided by
$2\pi$, $\Gamma_0$ is the intensity of the unit vortex quantum.

It may be supposed that the core of the superfluid vortex is a
cylindrical region of normal neutron fluid immersed in the
superfluid neutron sea. As a usual, the radius of the core of the
vertex, $a_0$, may be simply estimated as follows: in the core of
the vortex, the position uncertain of the neutron is $\Delta x\sim
a_0$, the momentum uncertain of the neutron is $\Delta p\sim h/a_0$
by the Heisenberg's principle of uncertainty and then the uncertain
for energy of the neutron is about $\Delta E\sim (\Delta
p)^2/2m_n\sim h^2/(2m_na_0^2)$. The neutron fluid in the core of the
vortex will be in the normal state (no Cooper pairs), where the
energy uncertain of the neutron is greater than the energy gap of
the neutron superfluid (or the binding energy for the Cooper pair of
neutrons) $\Delta E>\Delta_n$. Thus we have $a_0\approx
h/\sqrt{2m_n\Delta_n}$. For the vortexes of isotropic neutron
superfluid $\Delta(^1S_0)\approx 2$MeV, $a_0\sim10^{-12}$cm and for
the vortexes of isotropic neutron superfluid $\Delta(^3P_2)\approx
0.045$MeV, $a_0\sim10^{-11}$cm.

Outside the core of the vortex, neutrons are in a superfluid state.
The superfluid neutrons revolve round the vortex line with a
velocity \citep{b12}
\begin{equation}
 v_s(r)=\frac{n\hbar}{2m_nr},
 \label{2}
\end{equation}
where $r$ is the distance from the axis of the vortex filament. The
distribution of the angular velocity of neutron superfluid revolving
around the vortex filament is
\begin{equation}
 \omega_s(r)=\frac{n\hbar}{2m_nr^2},
 \label{3}
\end{equation}

Therefore the revolution of superfluid neutrons around the vertex
filament is placed in a differential state. Near $r\sim a_0$, the
angular velocity reaches the largest value is given by
\begin{equation}
  \omega_{s,\rm{max}}=\frac{n\hbar}{2m_na_0^2},
 \label{4}
\end{equation}
We have $\omega_{s,\rm{max}}(^1S_0)\sim 10^{21}n$ and
$\omega_{s,\rm{max}}(^3P_2)\sim 10^{19}n$. Inside the core of the
vertexes ($r<a_0$), however, the normal neutron fluid revolves
rigidly at angular velocity of $\omega_{s,\rm{max}}$.

According to \citep{b12}, the number of superfluid vertex filaments
per unit area is $2\Omega/n\Gamma_0$. Then the order of magnitude of
the average separation, $b$, between vertex filaments and the total
number, $N_{\rm{Vertice}}$, of the superfluid vertexes in the
superfluid region are respectively
\begin{equation}
 b=(\frac{\bar{n}\hbar}{2m_n\Omega})^{1/2},
 \label{5}
\end{equation}
\begin{equation}
 N_{\rm{Vertice}}=\frac{2m_n\Omega}{\bar{n}\hbar}R_s^2,
 \label{6}
\end{equation}
where $\Omega$ is the angular velocity of rotation as a whole, $R_s$
is the radius of the region of the superfluid. $\bar{n}$ is the
circulation quantum number of each vortex filament on the average.

It is generally believed that the circulation quantum number of each
vortex filament for liquid $^4$He and $^3$He  in the Earth's low
temperature is very low, even in the lowest basic state $n=1$ for
thermal dynamical equilibrium. However, the interior of the nascent
neutron star must be in a turbulent vortex state. This is because
that neutron stars originated from the collapsed supernovae core
during violent supernovae explosion in a very short time less than
10 seconds and it is hard to transport the rotational angular
momentum of the collapsed supernovae core outwards. It not only
rotates fast (but its angular velocity cannot exceed the critical
angular velocity to maintain neutron star stability) but also store
considerable part of the stellar angular momentum in the violent
turbulent state of the neutron fluid \citep{b40,b41}. The classical
circulation of vortex filaments intensity ($\Gamma$) may be  very
large. As the interior temperature of the neutron star is lowered to
below the critical temperature of the isotropic superfluid $^1S_0$
state and the anisotropic superfluid $^3P_2$ state respectively, the
$^1S_0$ state is isotropic while the $^3P_2$ state becomes
anisotropic. At this time, the violent classical turbulent vortex
state with very large vortex quantum number. We expect that for very
young pulsars, the quantum number n in Eq. (2) may reach above
$10^2-10^4$.

In the anisotropic superfluid region of neutron Cooper pairs. Every
$^3P_2$ Cooper pairs consist two neutrons with parallel spins so
that the net spin of the $^3P_2$ Cooper pair is equal to 1. Every
pair possesses abnormal (anomalous) magnetic moment with magnitude
twice as that of the neutron ($\mu_n$). The total number of neutrons
contained in the Cooper pairs in the $^3P_2$ anisotropic superfluid
in neutron stars is 8.7\% of the total of neutrons number in that
region \citep{b44}. At high temperatures with $\mu_nB/kT\ll1$, in
the presence of external magnetic fields, the direction of the
magnetic moments of the Cooper pairs in the anisotropic $^3P_2$
neutron superfluid region is very chaotic almost reaching the equal
probability state (ESP). The anisotropic superfluid state at this
time is called phase A. Here $\mu_n$ is the magnetic moment of
neutron, $T$ the interior temperature, $k$ is the Boltzmann constant
and $B$ is the external magnetic field strength.

When there exists possible elective cooling process in neutron
state, the temperature of the anisotropic $^3P_2$ superfluid region
may be lowered to below the Curie temperature ($\mu_nB/kT\gg1$) the
majority moments of the $^3P_2$ Cooper pairs are spontaneously
arranged in the same direction as the external strong magnetic field
(similar to the formation of magnetic domains in the low temperature
laboratories). This anisotropic superfluid state is called phase B.
The phase A and B in the anisotropic superfluid in neutron stars are
similar to the those of in the anisotropic superfluid $^3$He in low
temperature laboratories near absolute zero ($T<0.02$K). But the
phase B of the $^3P_2$ neutron superfluid possesses very strong
magnetic fields and the effective magnetic moments of the $^3P_2$
Cooper pairs are also very strong \citep{b44,b45}. According to the
theory proposed in our works \citep{b40,b41}, the magnetic moments
of the $^3P_2$ Cooper pairs can emit very strong magnetic dipole
radiation as the neutron Cooper rotate around the axis of the
superfluid vortex . This radiation may be considered as an effective
heating mechanism in neutron star interiors. Our present paper is
just based on this idea.

\section[]{Our researches on neutron stars and the cooling and heating problem in neutron star interiors}
\subsection[]{Our researches on neutron stars}

We proposed a theory in 1982 \citep{b41} that the neutrino radiation
by neutron superfluid vortexes of neutron stars is a decisive factor
for spin down of pulsars with longer period ($P>1.0$s) and the rate
of spin down, $\dot{P}$, is proportional to $P^2$($\dot{P}\propto
P^2$),\textbf{ which was repeatedly supported by statistical works
of pulsars from \citet{b370,b371,b372}}. \textbf{It is also
confirmed by the recent observed ($P-\dot{P}$) diagram of pulsars
(ATNF Pulsar Catalogue, 2016 see Fig.1)}, (\textbf{we can see the
link of http://www.atnf.csiro.au/research /pulsar/psrcat/}) and
which deviates seriously from the model of magnetic dipole radiation
with the relation $\dot{P}\propto P^{-1}$ (the standard model). The
observed pulsar ($P-\dot{P}$) diagram (ATNF Pulsar Catalogue, 2016)
is not only supporting the spin down mechanism by the neutrino
radiation from neutron superfluid vortexes of neutron stars, but
also is supporting the idea of existence for neutron superfluid
vortexes of neutron stars. Therefore, it is also in favor of our
another paper \citep{b18} in which we proposed a pulsar heating
mechanism by magnetic dipole radiation from the anisotropic $^3P_2$
neutron superfluid vortex (afterwards $^3P_2$ MDRA). In that paper,
however, we calculated only the heating rate in lower magnetic field
when $\mu_nB\ll kT$. In present paper, we will recalculate the
heating rate with two different case of lower magnetic field and
stronger magnetic field when $\mu_nB\gg kT$.

In the last decade we have studied the origin of the strong magnetic
fields of neutron stars and the origin of the super strong magnetic
fields of the magnetars \citep{b42,b43}. We have systematically
considered the properties of magnetars such as the physics of the
high X-ray luminosity in terms of principles and methods of
condensed matter physics \citep{b44,b45}. We have also investigated
weak interaction rates and neutrino energy loss in magnetars
\citep{b25,b26,b27,b28,b29,b30,b31,b32}.

The magnetic fields of most pulsars are $10^{11}-10^{13}$ Gauss with
typical magnetic field strength $10^{12}$ Gauss. The average
magnetic field of the sun is one gauss, and the upper half main
sequence stars of large mass do not have surface convection, their
magnetic fields are not very strong except the Ap stars. The
collapse of the central region of stars of large mass during
supernova explosion can only produce magnetic fields of
$10^{9}-10^{11}$ Gauss. In other words, the primary magnetic fields
(or the fossil magnetic fields) of neutron star produced by the
collapse of the central core during supernova explosion (because the
conservative of magnetic induction flux) cannot reach
$10^{12}-10^{13}$ Gauss. It is even more difficult to obtain
magnetic fields of strength $10^{14}-10^{15}$ Gauss, of the
magnetars discovered by astronomical observations during the last
decades. Some authors suggested that magnetars may still originals
from neutron stars due to the core collapse during supernova
explosion. i.e. there may already exist very strong magnetic fields
in massive stars even before the collapse of the central core. It
seems that there are no convincing observation evidence to support
this theory. How to produce the strong pulsar magnetic field? This
is a very interesting and important question to astrophysics and
astronomers. We make use the methods of the very famous theory in
condensed matter physics to study and discuss the origin of the
strong pulsar magnetic fields. Similar to the production of the
induction magnetic gas in metals in the presence of external
magnetic fields, we found that the Pauli Spin paramagnetic effect of
the extreme relativity degenerate electron gas in neutron star
interiors can amplify the fossil magnetic field $B^{(0)}$
($10^{9}-10^{11}$ Gauss) (which was originated from the central
collapse of supernova explosion) by a factor of 90. The resulting
magnetic field may thus reach ($10^{12}-10^{13}$) Gauss \citep{b43}.

About the physical origin of the super strong magnetic fields of the
magnetars, we have already discussed the Pauli spin paramagnetic
effect of the extreme relativistic degenerate electron gas
\citep{b43,b44}. The main idea is that the magnetic moment of the
$^3P_2$ Cooper pairs in the presence of the external strong magnetic
fields which are already amplified through the Pauli spin
paramagnetic effect, may also show some paramagnetic effect . This
is similar to the theory of magnetic domain in condensed matter
physics. When the interior temperature is lowered to below the Curie
temperature ($(1-2)\times10^7$K), the magnetic moments of most of
the $^3P_2$ Cooper pairs are spontaneously aligned in the same
direction and leading to super-strong magnetic fields of the
magnetic \citep{b44}. Of course, larger is the mass of the $^3P_2$
anisotropic superfluid, stronger the magnetic field.

Since the discovery of pulsars in 1967, almost all the theoretical
discussion concerning pulsar physics quoted the density of state for
a relativistic electron gas in strong magnetic fields from the
popular statistical mechanics text books by \citet{b20,b39}.
\citet{b6,b16,b21,b22}) discussed the state equation of electron gas
and the properties of magnetars in super-strong magnetic fields in
detail. However, their theoretical investigations are not able to
explain the astronomical observations such as the mechanism for the
production of high X-ray luminosity in magnetars.

We also discussed the properties of magnetars and the mechanism for
the production of high X-ray. Firstly, in the presence of super
strong magnetic fields ($B>B_{\rm{cr}}$,
$B_{\rm{cr}}=4.414\times10^{13}$G is Landau critical magnetic
field), the Fermi energy of the election gas increases with the
magnetic field strength \citep{b45}
\begin{equation}
 E_{\rm{F}}(e)=42.9(\frac{B}{B_{\rm{cr}}})^{1/4}~~\rm{MeV}.
 \label{7}
\end{equation}

From Eq. (7), we may have two important conclusions \citep{b45} as
follows. Firstly, the mechanism to generate high X-ray luminosity of
magnetar can be naturally explained. Secondly, direct Urca
(hereafter DUrca) process ($p+e^-\longrightarrow n+\nu_e,
n\longrightarrow p+e^-+\bar{\nu_e}$) may happen in super-strong
magnetic fields.

In the current model of neutron stars, there are 95\% neutrons, 5\%
protons and the number of electrons are equal to the number of
protons so as to maintain charge neutrality. The DUrca process is
forbidden in the $\beta$ equilibrium neutron star interior. This is
because the conservation of energy and momentum cannot be
simultaneously satisfied unless the fraction of protons is more than
9\% (It is might be possible in the inner core of neutron stars, but
it is most believed that the material in the inner core is really
made by quarks) \citep{b53,b38}.

However, it is totally different in the presence of super-strong
magnetic field. Really the Fermi energy of the relativistic
degenerate electron gas in the neutron star interiors increases with
the magnetic field in the super-strong magnetic field
$B>B_{\rm{cr}}$ \citep{b45}. The Fermi energy of the electrons is
apparently exceed the Fermi energy of the neutrons of the
non-relativistic degenerate neutron gas. Therefore the current
abstinence rule above is broken in the superstrong magnetic field.

\textbf{We note that the DUrca process in strong magnetic field had
been discussed in detail by \citet{b51}. In the abstract of their
paper, however, they declared that in the case of superstrong
magnetic fields, such that e and p populate only the lowest Landau
levels is briefly outlined. Their idea is not suitable for neutron
star research evidently. The reason is following:  Their idea is
just suitable for the Boltzmann's classical gas really.  But in the
neutron stars, the highly degenerate quantum charged Fermi gas (e.g.
e and p) in the superstrong magnetic fields, the filling of e and p
is totally different with ones for the Boltzmann 's classical gas
due to both the Pauli exclusion principle and the number of the
electrons in a unit volume is finite (See the Fig.1 and Fig.2 of
\citet{b45}). Besides, It is showed that the largest magnetic field
of magnetars is about $3\times10^{15}$ G \citep{b44}, which is
really a magnetic domain of the 3P2 neutron anisotropic superfluid
under Curie temperature. The case with higher magnetic fields
($\geq10^{18}$G) in the paper by \citet{b51} is only an assumption
without physical reason. Therefore, we discuss it further according
to our own idea regardless of their work.}

From this it can be concluded that the DUrca process is allowed in
the presence of super strong magnetic fields. The DUrca process
necessarily leads to the following two important consequences. (1)
The DUrca process can supply the most effective cooling mechanism to
the young pulsars. (2) In addition, the DUrca process can also
supply the effective cooling mechanism to the anisotropic $^3P_2$
superfluid region in neutron star interiors in super-strong magnetic
fields.

For magnetic fields not very strong we have proposed a pulsar
heating mechanism, magnetic dipole radiation emitted by the
magnetic, moments of the Cooper pairs with parallel spins in the
anisotropic $^3P_2$ superfluid vortex motion in neutron stars
\citep{b18,b40}. In this paper, we will generalize the heating
mechanism mentioned to the case of super strong magnetic fields. We
also combine the generalized heating mechanism (in super-strong
magnetic fields) with effective cooling mechanism in super-strong
magnetic fields (i.e., the DURCA process) to investigate the phase
oscillation between phase A and phase B in the anisotropic $^3P_2$
superfluid similar to the phase A and phase B in the liquid $^3$He
superfluid. It is expected that this phase oscillation may be the
desired mechanism to explain pulsar glitch. This is a theoretical
basis of this paper.

\subsection[]{The cooling and heating problem in neutron star interiors}

\subsubsection[]{The Cooling mechanism in neutron star interiors}

The cooling problem in neutron star interiors is an important and
difficult problem in the last fifty years. Neutron stars are lorn as
the remnant of supernova explosion. initial temperature of the
nascent neutron stars is roughly $10^{11}$ K. But the observed
surface temperature of young pulsars is roughly only $10^{6}$ K. For
example the Crab nebula pulron (PSR0531) is the remnant of the
supernova explosion during 1054 with surface temperature less than
$10^{6}$K. What is the cooling mechanism that cooled the nascent
neutron star very rapidly in less than a thousand year. The general
relativistic effect of gravitational radiation can be effective only
in the initial fear hours. This effect can cool the neutron star by
at most two orders of magnitude. In the two decades between 1970 and
1990, it is hoped that the so called $\pi-$condensation in nuclear
physics may be effective to cool the nascent neutron star. However,
the $\pi-$condensation has never been discovered in nuclear physics
experiment. Thus, the theoretical assumption of the
$\pi-$condensation is unreliable. On the other hand, if the direct
Urca process is possible in neutron stars, it is a very effective
cooling mechanism because energy is very rapidly carried away by the
emitted ($\nu_e, \bar{\nu_e}$) pairs. Thus, it is still generally
expected that the possible DUrca in neutron stars is the effective
cooling mechanism. The neutron energy loss rate (cooling rate) in
the DUrca process is
\begin{equation}
 \varepsilon_\nu^{\rm{DUrca}}\equiv E_\nu^{\rm{DUrca}}=10^{21}\Re T_8^6~~~\rm{ergs~cm^{-3}~s^{-1}},
 \label{8}
\end{equation}
where $\Re$ is the superfluid suppress factor.  For $T\ll
T_{\rm{cr}}$, $\Re\sim (T/T_{\rm{cr}})^2$, otherwise $\Re\sim
(T/T_{\rm{cr}})$. This superfluid suppress factor is not the usual
Boltzmann suppress factor $e^{-(T/T_{\rm{cr}})}$. $\Re
\approx(0.1\sim0.2)$ for $T_8=1.0$, and $\Re \approx(0.01\sim0.05)$
for $T_8=0.2$. As we noted before that the direct Urca (DUrca)
process is forbidden in normal neutron stars because conservation of
energy and momentum cannot be simultaneously satisfied \citep{b53}.
The neutrino emission rate (cooling rate) of the modified Urca
process with the suppress factor due to the neutron superfluid
($n+p+e^-\longrightarrow n+n+\nu_e$, or MUrca process) is
\begin{equation}
 E_{\nu}^{\rm{MUrca}}=7.4\times10^{12}(\frac{\rho}{\rho_{\rm{nuc}}})^{2/3}T_8^8e^{-T^{*}/T}~\rm{ergs~cm^{-3}~s^{-1}},
 \label{9}
\end{equation}
where $T^{*}=\Delta (^3P_2)/2k\approx 2.8\times10^8$K.

This is a high order weak interaction of six particles so it is a
much less effective process than the direct Urca process as a
cooling mechanism. Another important cooling mechanism is the
so-called PBF (pain breaking and formation) neutrino emission
mechanism. When the temperature of the neutron star reaches the
transition temperature so that normal neutrons transform to
superfluid neutrons (i.e., the two free neutrons with adverse
direction at the surface of the Fermi sphere form Cooper pairs), the
weak neutral current leads to the emission of neutrino pairs
$n+n\longrightarrow [n, n]+\nu \bar{\nu}$. Thermal agitation or
other heating mechanism can supply energy to break the Cooper pairs
. This cyclic process may then lead to neutron star cooling. The
neutrino energy emission rate (i.e., the cooling rate) is
\begin{equation}
  E_\nu^{\rm{PBF}}=A_{\rm{PBF}} T_8^7~~~\rm{ergs~cm^{-3}~s^{-1}},
 \label{10}
\end{equation}
where $A_{\rm{PBF}}=10^{15}$. This process can happen only during
the phase transition from normal neutrons to superfluid neutrons.
For superfluid without normal neutrons (for temperatures obviously
lower than the phase transition temperature) this PBF cooling
mechanism is ineffective.

The most difficult problem to overcome in the cooling of neutron
star interior is whether or not the direct Urca process is possible.
It is impossible in the current theory. This is because the core
region of neutron stars is in a quark state and outside the core the
proton component is about 5\%. According to the current theory the
direct Urce process is impossible to occur. However, this taboo is
broken by our recent paper in 2016 \citep{b45}. We pointed out that
in the presence of super-strong magnetic fields of $10^{15}$ Gauss
of the anisotropic superfluid $^3P_2$ state in the young pulsars
(and especially magnetars $B>B_{\rm{cr}}$), the direct Urca process
is possible. This is because the Fermi energy of the electron gas
increases with the magnetic field strength in the presence of
super-strong magnetic fields in magnetars (see Eq.(7)). The Fermi
energy of the relativistic degenerate electron gas exceeds the Fermi
energy (about 60 MeV) of non-relativistic degenerate neutron gas.
Thus, the protons near the non-relativistic degenerate proton Fermi
surface may join the electrons near the relativistic degenerate
electron Fermi surface to form neutrons $p+e^-\rightarrow n+\nu_e$;
the energy of the emitted neutrons exceed the Fermi energy of the
non-relativistic degenerate neutron gas. The emitted high energy
neutrons may emit electrons via beta decay $n\rightarrow
p+e^-+\bar{\nu}_e$. In this way the direct Urca process is possible.

The direct Urca process may then lead two important effects.
Firstly, it can provide an effective quick cooling mechanism for the
young pulsars. Secondly, the direct Urca process may lead to
effective cooling in the neutron anisotropic superfluid region. In
addition, we have also proposed that the heating mechanism due to
the magnetic dipole radiation of the magnetic moments of the
anisotropic $^3P_2$ superfluid vortex motion is very effective
\citep{b18,b40}. The combined effects of the cooling mechanism and
heating mechanism in the anisotropic $^3P_2$ neutron superfluid may
then lead to the phase oscillation between phase A and phase B in
neutron stars similar to the phase A and phase B in liquid
superfluid $^3$He. This phase oscillation can naturally explain
pulsar glitch. Due to the cooling rate of the DUrca process exceeds
the cooling rates of both modified Urca process and the PBF process
and the cooling rates of both the MUrca and the PBF process can be
neglected.

\textbf{Here we would like to point out that the magnetic field in
the region of the anisotropic neutron superfluid $^3P_2$  state will
arrive at above $10^{15}$ G. The reason is as follows: The observed
magnetic field on the polar region of young pulsars are about
$(2-5)\times10^{12}$ G usually. The anisotropic neutron superfluid
$^3P_2$ state is about in the region 2Km$<r(^3P_2)<$5Km . Due to
dipole magnetic fields decrease with $B(r)\propto r^{-3}$, we may
suppose that the strength of the magnetic field in the region of the
anisotropic neutron superfluid 3P2 state will arrive at above
$10^{15}$ G. Therefore we may take the DUrca process as the
effective cooling in the neutron anisotropic superfluid region.}

\subsubsection[]{The heating mechanism in neutron star interiors}

\textbf{We have proposed that the heating mechanism due to the
magnetic dipole radiation of the magnetic moments of the anisotropic
$^3P_2$ superfluid vortex motion is very effective \citep{b18,b40}.
The combined effects of the cooling mechanism and heating mechanism
in the anisotropic $^3P_2$ neutron superbluid may then lead to the
phase oscillation between phase A and phase B in neutron stars
similar to the phase A and phase B in liquid superbluid $^3$He. This
phase oscillation can naturally explain pulsar glitch.}

We reinvestigted and improved our early model above \citep{b18,b40}.
The magnetic dipole radiation of the anisotropic neutron superfluid
in neutron stars is recalculated. We now elaborate in more detail
this heating mechanism. The magnetic moments of the Cooper pains in
the anisotropic $^3P_2$ neutron superfluid will produce magnetic
dipole radiation as the Cooper pairs rotate around the axis of the
superfluid vortex. In the strong external magnetic field, the number
of the $^3P_2$ Cooper pairs with magnetic moments anti-parallel to
the external magnetic field is more than the number of the $^3P_2$
Cooper pairs with magnetic moments parallel to the external magnetic
field. This leads to the parallel magnetic moments as the
temperature is lowered. In particular, magnetic moments are
strengthened at low temperatures. These superfluid vortex neutrons
rotate with high angular velocity around the superfluid axis. the
closer the neutrons from the vortex axis the larger the angular
velocity (inversely proportional to the square of the distance from
the vortex axis, the highest angular velocity may reach above
$10^{20} \rm{s}^{-1}$). We note that the magnetic dipole radiation
from the magnetic moments of the $^3P_2$ Cooper pairs is operating
via the Ekman pump cycle \citep{b18}. This would lead to the motion
of the normal neutrons toward the deep interior in neutron stars.
The thermal X-ray photons due to the magnetic dipole radiation from
the magnetic moments of the $^3P_2$ Cooper pairs are quickly
absorbed by matter because of the high opacity in the stellar
interior and then become a heating mechanism.

In our previous work \citep{b18,b40}, we actually discussed only the
case for magnetic fields not very strong and at relatively high
temperatures. We are now considering  superstrong magnetic fields
($B>B_{\rm{cr}}$). In this case the anomalous moments of the
anisotropic $^3P_2$ superfluid neutron Cooper pairs tend
spontaneously to become anti-parallel to external magnetic field
(and thermal agitation tend to make the direction of the magnetic
moments chaotic). This is similar to the magnetic moments of the
tries election in metals tend to become the Pauli paramagnetism and
can produce induction magnetic moments. The total induction magnetic
moments of  the $^3P_2$  neutron superfluid is \citep{b44}
\begin{equation}
  \mu_{\rm{pair}}^{\rm{tot}}(^3P_2)=\mu_nqN_{\rm{A}}m(^3P_2)f(\mu_nB/kT),
 \label{11}
\end{equation}

The average effective magnetic moment of each neutron Cooper pain in
$^3P_2$ neutron superfluid is
\begin{equation}
  \bar{\mu}_{\rm{n}}^{\rm{eff}}=\mu_nqN_{\rm{A}}f(\mu_nB/kT),
 \label{12}
\end{equation}
\begin{equation}
  q=3(\frac{\Delta(^3P_2(n))}{E_{\rm{F}}(n)})^{1/2}\approx 8.7\%,
 \label{13}
\end{equation}
Where $q$ is the fraction of the neutrons that combined into the
$^3P_2$ Cooper pairs. $\Delta(^3P_2(n))$ is the energy gap of the
superfluid. $f(\mu_nB/kT)$ is the Brillouin function.
\begin{equation}
  f(x)=\frac{2\sinh(2x)}{1+2\cosh(2x)},
 \label{14}
\end{equation}
when $x\ll1$, $f(x)\approx 4x/3$, otherwise, $f(x)\rightarrow1$.

When the magnetic moment of the Cooper pair rotate around the axis
with angular velocity, the power emitted by magnetic dipole
radiation is
\begin{equation}
  w(n)=\frac{2\omega^4}{3c^2}\bar{\mu}^2_{\rm{eff}}\sin^2 \alpha,
 \label{15}
\end{equation}
where $\alpha$ is the angle between the direction of the magnetic
moment and the direction of the axis of  rotation.

The axis of every superfluid vortex is parallel to the axis of
rotation of the neutron star , and the length of the vortex can be
approximately attained as the radius $R_P$ of the $^3P_2$ superfluid
region. The angular velocity of the $^3P_2$ Cooper pair at the
distance $r$ from the axis is $\omega(r)$, and the wavelength of the
emitted magnetic dipole radiation from the magnetic moment is
$\lambda(r)=2¦Ðc/\omega(r)$. Consider now a series of cylindrical
region along the vortex axis with height $\eta \lambda(r)$ ($\eta
\approx 1\%$) and radius $r\rightarrow r+dr$. The Cooper paint
rotate around the superfluid vortex axis with the angular velocity
$\omega(r)=nh/4\pi m_nr^2$. The magnetic dipole radiation emitted by
these Cooper paint all has wavelength $\lambda(r)$. The phases of
the magnetic dipole cyclotron radiation emitted by the rotating
magnetic moments of the $^3P_2$ neutron Cooper pairs around the
vortex filament axis are very close in this small cylinder.
Therefore, the amplitudes of these radiation are added. The magnetic
dipole cyclotron radiation intensity is proportional to the square
of the number of Cooper paint in this region, i.e., these
electromagnetic radiations are interfering with each other. The
power of the magnetic dipole radiation emitted by the neutron
superfluid in these interfering region is
\begin{equation}
  W(n)=\frac{2\omega^4}{3c^2}M^2_{\Delta{\rm{V}}}\sin^2 \alpha,
 \label{16}
\end{equation}
where $M_{\Delta{\rm{V}}}$ denotes the total magnetic moment in
$\Delta{\rm{V}}$
\begin{equation}
  M_{\Delta{\rm{V}}}=dN_{\Delta{\rm{V}}}\bar{\mu}_n^{{\rm{eff}}},
 \label{17}
\end{equation}

\begin{equation}
  dN_{\Delta{\rm{V}}}=\frac{\rho_n(r)}{m_n}\eta \lambda_s(r)2\pi rdr,
 \label{18}
\end{equation}
\begin{equation}
  \lambda_s(r)=\frac{2\pi c}{\omega_s(r)}=\frac{4\pi cm_nr^2}{n\hbar}~~~(r>a_0),
 \label{19}
\end{equation}
where $dN_{\Delta{\rm{V}}}$ are the number of neutron in
$\Delta{\rm{V}}$, and $\lambda_s (r)$ is the magnetic dipole
radiation photon wavelengths from the vortex motion of $^3P_2$
neutron superfluid. $a_0$ is the core radius of the $^3P_2$
superfluid vortex with normal neutrons, $a_0\sim (10-100)fm$.

The radiation power of one $^3P_2$ superfluid vortex is
\begin{eqnarray}
W_1^s&=\int_{a_0}^b\int_{a_0}^b\frac{R_p\sin^2\alpha}{\eta
\lambda_s(r)}\frac{2\omega_s^4}{3c^2}[\bar{\mu}_n^{\rm{eff}}]^2
\times \delta(\omega_s-\frac{n\hbar}{2m_nr^2})dN^2_{\Delta{V}}\nonumber\\
&=A\int_1^{b/a_0}\int_1^{b/a_0}\omega^{3}(r^{'})r^{'2}\delta(\omega-r^{'-2})(dr^{'})^2,
 \label{20}
\end{eqnarray}
where
\begin{eqnarray}
A&=\frac{2\pi^3}{3c^2}(\frac{n\hbar}{m_n})^3(\frac{\rho_n}{m_n})^2
\times R_p\eta a_0^{-2}[\mu_nqf(\mu_nB/kT)]^2\overline{\sin^2\alpha}\nonumber\\
&=10^{24}(\frac{\bar{n}^3}{10^3})\frac{\eta}{0.01}(\frac{a_0}{10fm})^{-2}(\frac{\overline{\sin^2\alpha}}{0.1})\nonumber\\
&\times R_{p,5}[f(\mu_nB/kT)]^2~~~\rm{ergs~s^{-1}},
 \label{21}
\end{eqnarray}
where $R_{p,5}\equiv R(^3P_2)/(1\rm{Km})$. Making use of Eq.(6), the
total power of the emitted magnetic dipole radiation from all the
$^3P_2$ anisotropic superfluid vortex is given by
\begin{eqnarray}
W=\frac{1}{3}N_{\rm{Vertice}}=AR_p^2\frac{m_n\Omega}{\bar{n}\hbar}(\frac{\overline{\sin^2\alpha}}{0.1})\nonumber\\
=2.0\times10^{39}R^3_{p,5}(\frac{\overline{n^3}}{10^2\overline{n}})\frac{\eta}{0.01}(\frac{a_0}{10fm})^{-2}\nonumber\\
\times(\frac{P_{\rm{SF}}(^3P_2)}{1ms})^{-1}[f(\mu_nB/kT)]^2~\rm{ergs~s^{-1}}.
 \label{22}
\end{eqnarray}

The heating rate (i. e., the radiation power per unit volume) due to
the magnetic dipole radiation from the $^3P_2$ neutron superfluid
vortex in neutron star is
\begin{eqnarray}
\varepsilon\approx2.0\times10^{24}(\frac{\overline{n^2}}{10^3\overline{n}})\frac{\eta}{0.01}(\frac{a_0}{10fm})^{-2}(\frac{\overline{\sin^2\alpha}}{0.1})\nonumber\\
\times(\frac{P_{\rm{SF}}(^3P_2)}{1ms})^{-1}[f(\mu_nB/kT)]^2~\rm{ergs~cm^{-3}~s^{-1}},
 \label{23}
\end{eqnarray}

The heating power of the dipole radiation from the $^3P_2$ neutron
superfluid vortex in phase B, with $\mu_nB/kT\gg1$ is ($^3P_2$MDRA )
\begin{eqnarray}
\varepsilon^{(\rm{B})}\approx2.0\times10^{24}(\frac{\overline{n^2}}{10^3\overline{n}})\frac{\eta}{0.01}(\frac{a_0}{10fm})^{-2}(\frac{\overline{\sin^2\alpha}}{0.1})\nonumber\\
\times(\frac{P_{\rm{SF}}(^3P_2)}{1ms})^{-1}~\rm{ergs~cm^{-3}~~s^{-1}},
 \label{24}
\end{eqnarray}

The heating rate for phase A with $\mu_nB/kT\ll1$ is the same as our
results ($^3P_2$MDRA) in 1982 (e.g., \citet{b18})
\begin{eqnarray}
\varepsilon^{(\rm{A})}\approx2.0\times10^{18}(\frac{\overline{n^2}}{10^3\overline{n}})\frac{\eta}{0.01}(\frac{a_0}{10fm})^{-2}(\frac{\overline{\sin^2\alpha}}{0.1})\nonumber\\
\times(\frac{P_{\rm{SF}}(^3P_2)}{1ms})^{-1}(\frac{B_{12}}{T_8})^2~~\rm{ergs~cm^{-3}~s^{-1}},
 \label{25}
\end{eqnarray}
where $B_{12}=B/10^{12}$Gauss, $T_8=T/10^8$K.

When we compare Eq.(24) and Eq.(25) with Eq.(8) for the cooling
mechanism DUrca, we obtain
\begin{equation}
 \varepsilon^{(\rm{B})}\gg\varepsilon^{(\rm{DUrca})}\gg\varepsilon^{(\rm{A})}.
 \label{26}
\end{equation}

\begin{figure*}
\centering
\includegraphics[width=10cm,height=10cm]{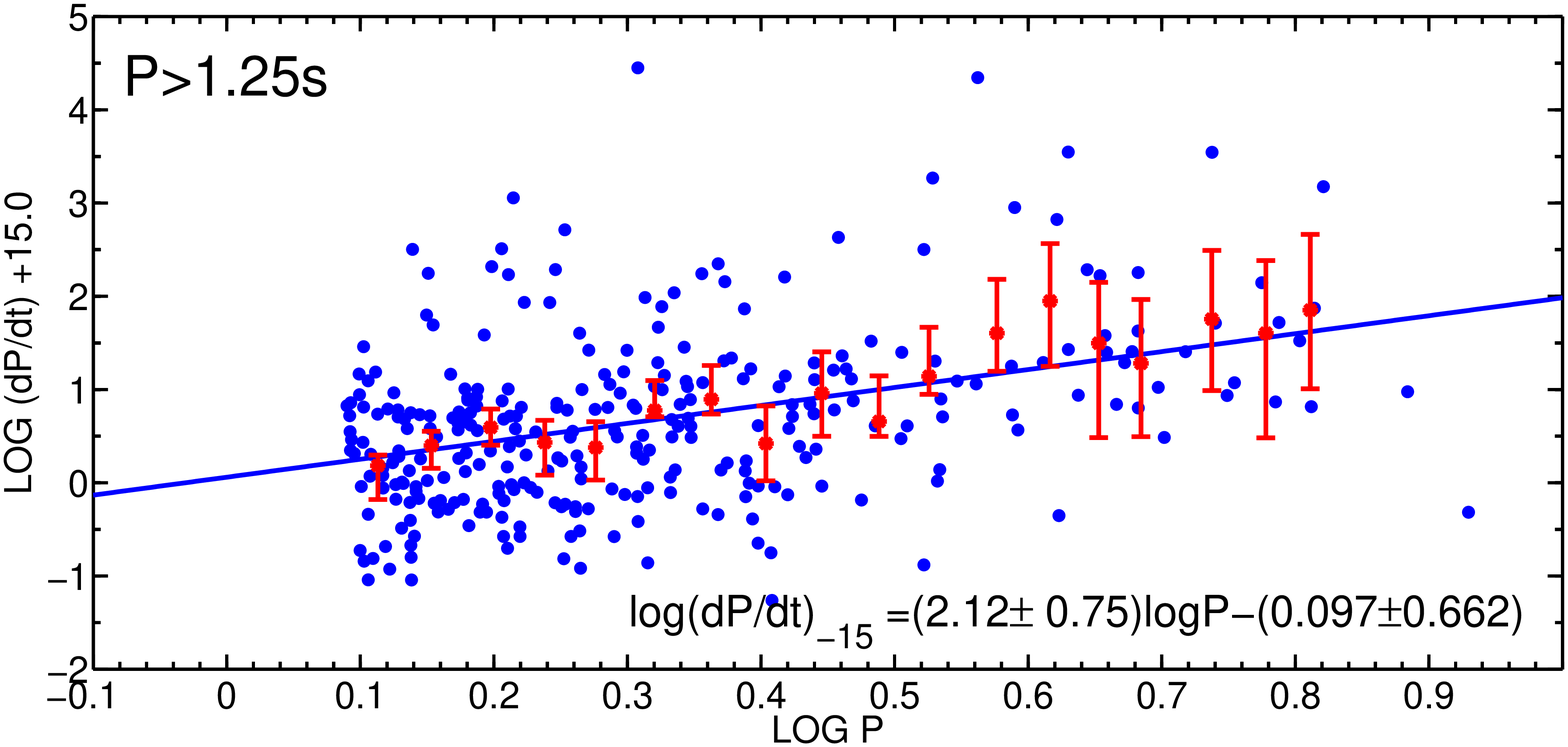}
  \caption{The observed ($P-\dot{P}$) diagram of pulsars with period longer than $P>1.25$ s (ATNF Pulsar Catalogue, 2016)
http://www.atnf.csiro.au/research/pulsar/psrcat/.\label{fig1}}
\end{figure*}

\begin{figure*}
\centering
    \includegraphics[width=15cm,height=15cm]{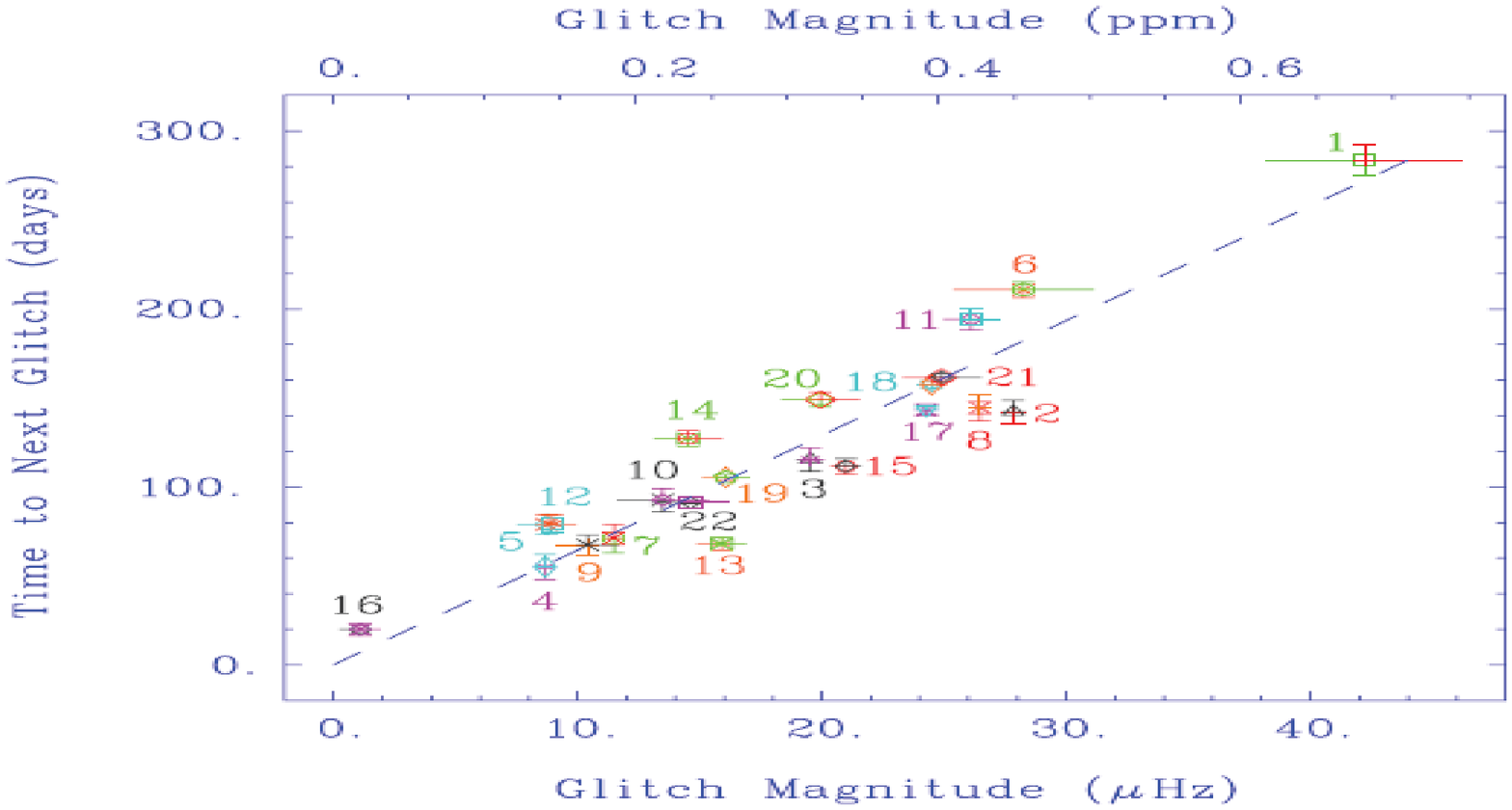}

\caption{The relationship between Glitch amplitude and the
stationary time interval for the young pulsar PSR J0537-6910. The
time bounds drawn for the points are equal (up and down) and each is
a quarter of the sum of two time intervals bounding the data
segment.  The slope of the dashed line fitted to the points and
through the origin is 6.4394 days $\mu$Hz$^{-1}$, or 399.37 days
ppm$^{-1}$\citep{b36}. \label{fig2}}
\end{figure*}

\section[]{Our model and the properties pulsar glitch}

In 2006, we proposed a phase oscillation model between normal
neutron fluid and $^3P_2$ neutron superfluid vortex state
\citep{b42} to explain the sudden change of pulsar periods (Glitch)
which is a difficult puzzle in pulsar physics. Although this model
can be used to explain the grand glitches of very young pulsars but
because there are too many undetermined free parameters in the
model, the important physical quantities and time scales are very
difficult to estimate, thus, these are the serious defect of the
model.

The main objective is to extend our 2006 model to a new model that
can explain the more general aspect of the observed pulsar glitch
phenomena on the basis of Eq. (26). We propose a new oscillation
model between phase A and phase B to reinterpret the pulsar glitch
phenomena

\subsection[]{The phase oscillation of the anisotropic $^3P_2$ neutron superfluid in neutron stars}

When the temperature in neutron stars in lowered to below the Curie
temperature ($\mu_nB/kT\gg1$), the majority of the $^3P_2$ neutron
Cooper pairs tend spontaneously to orient against the external
magnetic field (It is noted that the neutrons and $^3P_2$ neutron
Cooper pairs have abnormal magnetic moments). This phase B, $^3P_2$
neutron superfluid state has very strong magnetic field and the
effective magnetic moments of the $^3P_2$ neutron Cooper pairs are
rather strong. The magnetic moment of Cooper pairs will generate
very strong magnet dipole radiation with the heating rate
$\varepsilon^{\rm{B}}$ (see Eq.(24)) as the $^3P_2$ Cooper pairs
rotate around the superfluid vortex axis. As we noted before, this
is an effective heating mechanism in neutron stars. When the heating
rates of this mechanism exceeds the cooling rates of the possible
cooling mechanism exist in neutron stars (such as the DUrca
process), the temperature in neutron stars would rise very quickly,
the resulting thermal agitation would destroy the orderly
arrangement of the magnetic moments of the $^3P_2$ neutron Cooper
pairs. Once the temperature rises to $\mu_nB/kT\ll1$, the magnetic
moments of the $^3P_2$ Cooper pairs become completely chaotic, the
$^3P_2$ superfluid then transforms back phase A (ESP state, i.e.,
equal probability state). When the $^3P_2$ neutron superfluid
transforms to phase A, the strong induced magnetic moments of phase
B would also disappear. For $^3P_2$ neutron superfluid in phase A,
the effective magnetic moment of the $^3P_2$ Cooper pairs are very
weak, and the resulting magnetic dipole radiation emitted by these
$^3P_2$ Cooper pairs are also very weak (see $\varepsilon^{\rm{A}}$
from Eq.(25)).  From Eq.(26), it can be seen that the heating rate
of this heating mechanism in the phase A is far less than the
cooling rate of the cooling process in neutron stars (mainly the
DUrca process). The temperature in neutron stars would gradually be
lowered. When the thermal energy $kT$ is lowered to below the energy
of the magnetic moments of the $^3P_2$ Cooper pairs
($\mu_nB/kT\gg1$), then, the magnetic moments of the $^3P_2$ Cooper
pairs once again to spontaneously arrange themselves anti-parallel
to the external magnetic field, and phase B is recovered. According
to Eq.(26), the resulting magnetic dipole radiation is very strong.
The heating rate or this heating mechanism is much higher than the
cooling rate of the DUrca process. This is the formation of the
phase oscillation between phase A and phase B in the $^3P_2$ neutron
superfluids.

During this heating process with the heating rate
$\varepsilon^{\rm{B}}$, the thermal energy supplied from the
$^3P_2$MDRA would perturb the arrangement of the direction of the
magnetic moments of the $^3P_2$ Cooper pairs, and making them
chaotic. Besides, the thermal agitation due to the $^3P_2$MDRA
causes some parts (denoted by $\zeta$) of the $^3P_2$ neutron Cooper
pairs to break. Every broken $^3P_2$ Cooper pairs release two normal
neutrons. When the number of such released normal neutrons increases
to certain extent, the slowly rotating crust will be driven by the
much more fast rotation of the core $^3P_2$ superfluid due to the
strong coupling (by the nuclear force) of such normal neutrons with
normal protons in the anisotropic superfluid region. It is mentioned
in Section 1.2 that the neutron superfluid vortices may not coexist
with the proton super-conducting state \citep{b23} and through
electromagnetic coupling of protons in the $^3P_2$ superfluid core
with electrons in the crust of the neutron star. We know that the
electromagnetic interaction is an interaction through long distance.
This process thus generates the pulses glitch. Once the $^3P_2$
superfluid in phase B transforms to phase A, the heating mechanism
disappears. The cooling mechanism then quickly cause the neutron
superfluid to deviate from the ESP state once more and phase B is
recovered from phase A. The induction magnetic moment of the $^3P_2$
neutron superfluid reappears again and generating very strong
induction magnetic fields.

\subsection[]{Pulsar Glitch generated by phase oscillation of the $^3P_2$
neutrons superfluid phase A and phase B}

\subsubsection[]{Phase oscillation of the $^3P_2$ neutrons superfluid
phase A and phase B}

The temperature of the $^3P_2$ neutron superfluid is conspicuously
lower than the phase transition temperature
($T_{\rm{cr}}=2.8\times10^8$K). When the $^3P_2$ neutron superfluid
is in phase B, there are no normal neutrons except in the region of
the superfluid vortex core. During the phase transition from phase B
to phase A, it is mainly the competing processes between the cooling
mechanism of the DUrca processes for neutrino emission and the
heating mechanism due to the magnetic dipole radiation from the
$^3P_2$ neutron superfluid vortex motion. Since the heating rate due
to the $^3P_2$MDRA from the magnetic moments of the $^3P_2$ neutron
cooper pairs of phase B is much more than the cooling rate by the
DUrca process
($\varepsilon^{(\rm{B})}\gg\varepsilon^{(\rm{DUrca})}$), thermal
energy is supplied to the system. This causes the directions of the
magnetic moments of the $^3P_2$ neutron Cooper pairs to gradually
become completely chaotic and recover phase A. This period of time
is the heating time scale.

For the phase oscillation between phase A and phase B , once the
$^3P_2$ neutron superfluid phase B transformed into phase A, the
original strong heating rate ($\varepsilon^{(\rm{B})}$) of phase B (
B-$^3P_2$MDRA) becomes very weak ($\varepsilon^{(\rm{A})}$) of phase
A (A-$^3P_2$MDRA), the DUcar cooling mechanism dominates and the
$^3P_2$ neutron superfluid deviates again from the ESP state after a
period of the cooling time scale. The $^3P_2$ neutron superfluid
phase A transformed to phase B again. The $^3P_2$ neutron superfluid
again induce magnetic moments and generate corresponding induction
magnetic fields. This is the oscillation between phase A and phase
B.

\subsubsection[]{The release of normal neutron}

As we mentioned before, the thermal agitation due to the $^3P_2$MDRA
causes some parts ($\zeta$) of the $^3P_2$ neutron Cooper pairs to
break during the heating period. That means during the heating
period some nasal neutrons frozen in the Cooper pairs with fraction
$\zeta$ are released to become normal neutrons during the heating
period. In other words, at the same time as the $^3P_2$ neutron
superfluity transformed back to phase A, there are $\zeta N(^3P_2)$
neutron Cooper pairs are broken to become normal neutrons (total
numbers are $2\zeta N(^3P_2)$). The fraction of the normal neutrons
released is $\zeta q$ (where the ratio of the number of neutrons in
all the Cooper pairs to the total number of neutrons is $q$).

\subsubsection[]{The appearance of glitch}

\textbf{Although the type II superconducting protons (with magnetic
tubes) might appear \citep{b191}, but based on the arguments of
\citet{b23} that neutron superfluous vortex region may be no coexist
with the type II superconducting region, we may regard that the
protons are normal Fermi fluids in the neutron superfluid vortex
region.} The protons are tightly coupled to the elections in neutron
star interiors via coulomb interaction and they are rotation with
the observed pulsar angular velocity ($\Omega=2\pi/P$) around the
axis of rotation of the neutron stars. The strongly interacting
protons with the elections are basically decoupled from the neutron
superfluid vortex region in neutron stars. The protons can only
interact with the small amount of normal neutrons in the neutron
superfluid vortex core via nuclear interaction, while the election
can only interact via the very weak electron magnetic moment
interactions.

But the number, $\zeta q$, of the normal neutrons released from the
broken neutron superfluid Cooper pairs due to heating process can
strongly coupled to the normal protons via strong nuclear
interaction. This strong coupling can cause the fast rotation of the
neutron superfluid core to drive the slowly rotating outer crust,
such that the rotation of the whole magnetosphere including the
outer crust to suddenly rotate much faster, leading to the
appearance of glitch. In other words, glitch is the result of the
sudden increase of rotation velocity of the slowly rotating outer
crust driven by the fast rotation of the neutron superfluid core.
There are suddenly changes during glitch. For instance, pulsar
periods are suddenly changed. The angular momentum are transported
to the outer crust. The neutron superfluid phase B immediately
transformed to phase A. Besides, the superfluity vertex quantum
number $n$ is apparently lowered.

\subsubsection[]{The time interval between successive glitches and cooling time scale}

Once $^3P_2$ neutron superfluid recover the ESP state (the
directions of the magnetic moments are chaotic) the induction
magnetic moments of phase B would disappear.  Phase B transformed to
phase A in the heating time scale $t_{\rm{heat}}$. The strong
heating rate ($\varepsilon^{(\rm{B})}$) of phase B, $^3P_2$MDRA
transformed to the very weak heating rate ($\varepsilon^{(\rm{A})}$)
of phase A, $^3P_2$MDRA and then the direct Urca process dominates
the cooling process (Eq.(26)). Some of the normal neutrons, $\zeta
N(^3P_2)$ released by the broken Cooper pairs due to heating are
transformed back to neutron Cooper pairs again. In the presence of
superstrong magnetic fields, the magnetic moments spin of some of
the Cooper pairs tend to spontaneously parallel to the external
magnetic field and deviate from the random ESP state. The system
tends again to the phase B at statistical equilibrium. In other
words, the ESP state of the $^3P_2$ neutron superfluid in phase A
completely transform to state phase B (that deviates completely from
the ESP state). Consequently, the time interval between successive
glitches is about heating time scale plus cooling time scale .

As we mentioned before, the normal neutrons from the broken neutron
Cooper pairs due to heating may recover to become the superfluid
neutron Cooper pairs again through the DUrca neutrino emission
processes. This leads to the decoupling of the neutron superfluid
interior from the outer crust again. The time required for
decoupling is determined by the cooling time scale.

\subsubsection[]{Repeated glitches of young pulsars $^3P_2$NSV phase A and phase B repeated oscillations}

When the superfluid transformed from phase B to phase A (the ESP
state), the induced magnetic moments become very weak as the state
phase B disappears ($
\varepsilon^{(\rm{B})}\gg\varepsilon^{(\rm{DUrca})}\gg\varepsilon^{(\rm{A})}$).
The heating rate $\varepsilon^{(\rm{A})}$ of the $^3P_2$MDRA is far
lower than the cooling rate of the DUrca, the temperature of the
$^3P_2$ neutron superfluid is lowered. When the temperature lowered
to below the Curie comparative, the majority of the magnetic moments
of the $^3P_2$ neutron Cooper pairs tend to spontaneously parallel
magnetic moments reappear. The system returns to phase B again. This
is of course just the phase oscillation between phase and phase B.
The phase oscillation just mentioned may happen repeatedly many
times and this is indeed the mechanism for repeated glitches
(quasi-periodicity). The time interval, $\Delta t_{\rm{persist}}$,
for the appearance of normal neutron fluids intermediate between
phase A and B denoted persist generally very short and is very
difficult to monitor. Exact calculation of $\Delta t_{\rm{persist}}$
is very complicate because the relevant physics invoiced. But the
duration of $\Delta t_{\rm{persist}}$ is very important for the
conservation of different pulsars. We may consider that $\Delta
t_{\rm{persist}}\approx t_{\rm{heat}}$(the heating time scale).

\subsubsection[]{The disappearance of pulsar glitch}

After repeated glitches the superfluous vortex quantum number is
gradually lowered and the heating rate of the $^3P_2$MDRA also
gradually lowered (see the Eq.(24)). The time interval between
successive glitches becomes gradually longer and the amplitude tends
to decrease. There is no definite relationship between the amplitude
and the time interval of the glitches by some unknown random chance.
Once the heating rate $\varepsilon^{(\rm{B})}$ is lowered to felon
the cooling rate $\varepsilon^{(\rm{DUrca})}$ of neutron stars, the
neutron $^3P_2$ Cooper pairs can no longer be broken and then the
$^3P_2$ neutron superfluity can no longer recover the normal neutron
fluid state mentioned before the phase  oscillation between phase A
and phase B is immediately stopped. This leads to the disappearance
of the glitches of the old pulsars. factually, the observation
evidence indicates that no glitch was observed for pulsars with
periods $P>0.7$s \citep{b34}. The prediction of the our model is
consistent with observation.

\subsection[]{The estimates of the relevant time scales}
\subsubsection[]{Heating time scale and duration time scale of glitch}

During the heating process, the origin orderly arrangement of the
magnetic moments of the $^3P_2$ neutron Cooper pairs becomes
completely random due to the  $^3P_2$ neutron Cooper pairs with the
fraction $\zeta$ absorbed the heat energy which is supplied by
$\varepsilon^{(\rm{B})}$ of the $^3P_2$MDRA from phase B. This heat
energy is very much larger than the cooling rate
$\varepsilon^{(\rm{DUrca})}$ of neutron stars. The heat energy
required during the heating process is
\begin{equation}
 Q=\zeta \cdot \Delta N_{\mp}\cdot 2\mu_nB,
 \label{27}
\end{equation}
where $\Delta N_{\mp}$ is The difference of the number density of
$^3P_2$ neutron Cooper pairs with paramagnetic and diamagnetic
moment \citep{b44}
\begin{equation}
 \Delta
 N_{\mp}=n(^3P_2-pair)f(\mu_nB/kT)=(q/2)N_{\rm{A}}f(\mu_nB/kT).
 \label{28}
\end{equation}

Thus we have
\begin{equation}
 Q=\zeta qN_{\rm{A}}\mu_nBm(^3P_2)\approx
 4.0\times10^{38}(\frac{\zeta}{10^{-8}})(\frac{m(^3P_2)}{0.1M_\bigodot})B_{15}~\rm{ergs}
 \label{29}
\end{equation}
\begin{eqnarray}
t_{(\rm{heat})}&=&\frac{Q}{(4\pi/3)R_P^3\varepsilon^{(\rm{B})}}\nonumber\\
&&\approx0.2(\frac{\overline{n^3}}{10^2\overline{n}})(\frac{\zeta}{10^{-8}})(\frac{\overline{\sin^2\alpha}}{0.1})(\frac{P_{\rm{SF}}(^3P_2)}{1ms})^{-1}\nonumber\\
&&\times(\frac{m(^3P_2)}{0.1M_\bigodot})B_{15}~~\rm{s},
 \label{30}
\end{eqnarray}

The heating time scale is also the characteristic time scale for the
growth of the number of normal neutrons in the $^3P_2$ neutron
superfluous region. It is also the duration time scale of the glitch
\begin{eqnarray}
\Delta t_{\rm{persist}}&\approx t_{(\rm{heat})}\approx0.2(\frac{\overline{n^3}}{10^2\overline{n}})(\frac{\zeta}{10^{-8}})(\frac{\overline{\sin^2\alpha}}{0.1})(\frac{P_{\rm{SF}}(^3P_2)}{1ms})^{-1}\nonumber\\
&\times(\frac{m(^3P_2)}{0.1M_\bigodot})B_{15}~~\rm{s},
 \label{31}
\end{eqnarray}
For young pulsars, the superfluous vortex quantify number is still
very high, and $\overline{n^3}/\overline{n}\gg (10^2-10^4)$. Thus,
$\Delta t_{\rm{persist}}$ is very small and then it is difficult to
discover.

\subsubsection[]{The cooling time scale of glitch}

During the glitch , the phase B of the $^3P_2$ neutron superfluity
immediately transformed to phase A. The heating rate
$\varepsilon^{(\rm{A})}$ is much lower than the cooling rate of the
direct Urea process. During the periled of the phase transition from
phase B to phase A, only a fraction $\zeta$ of the $^3P_2$ neutron
Cooper pairs are broken to become normal neutrons. Therefore, during
the cooling process, all the normal neutrons can form $^3P_2$ cooper
pairs again, the neutron superfluous recourse phase B again. The
time scale required for thin transformation is the cooling time
scale, which can he estimated as follows:

The total energy released by the broken fraction $\zeta$ of the
$^3P_2$ neutron Cooper pairs to become normal is $Q^{'}=\zeta
qN_{\rm{A}}\Delta_{^3P_2}\cdot m(^3P_2)$. This energy is lost at the
cooling rate $\varepsilon_\nu^{(\rm{Urca})}$. The cooling time scale
is then
\begin{eqnarray}
t_{\rm{cool}}&=&\frac{Q^{'}}{(4\pi/3)R_P^3\varepsilon^{(\rm{DUrca})}}\nonumber\\
&\approx&4.6\times10^6(\frac{\Re}{0.01})(\frac{\zeta}{10^{-8}})(\frac{R_p}{5\rm{Km}})^{3}(\frac{m(^3P_2)}{0.01})~~\rm{s},\nonumber\\
 \label{32}
\end{eqnarray}
Since $t_{\rm{heat}}\ll t_{\rm{cool}}$, $\Delta t_{\rm{interval}}
=t_{\rm{heat}}+t_{\rm{cool}}\approx t_{\rm{cool}}$. The time
interval between successive glitches is same as Eq.(32), and the
time scale is about a month.

\subsubsection[]{The amplitude of glitch ($\Delta\Omega/\Omega$) and the correlation of the glitch duration with amplitude}

The angular momentum of neutron star crust is
$J_{\rm{crust}}=I_{\rm{crust}}\Omega$, and the angular momentum of
the superfluid core of the neutron star is
$J_{\rm{core}}=I_{\rm{core}}\Omega_{\rm{core}}$. The change of the
core angular momentum during glitch is not only proportional to the
angular momentum of the neutron star superfluous core and the number
of normal neutrons in the superfluity core $\zeta
qN_{\rm{A}}m(^3P_2)$, but also proportional to the heating rate
($\varepsilon_\nu^{(\rm{B})}(R(^3P_2))^3$) of $^3P_2$MDRA heating
process of phase B, where the factor
$\varepsilon_\nu^{(\rm{B})}(R(^3P_2))^3$ represents the efficiency
of the glitch and $\varepsilon_\nu^{(\rm{B})}$ is given by Eq.(24).
The change of the pulsar angular frequency is then proportional to
the following factors,
\begin{equation}
  \frac{\Delta\Omega}{\Omega}\propto\frac{I_{\rm{core}}}{I_{\rm{crust}}}\cdot\frac{\Omega_{\rm{core}}}{\Omega}\zeta
qm(^3P_2)\varepsilon_\nu^{(\rm{B})}(R(^3P_2))^3.
 \label{33}
\end{equation}

It follows that for old pulsars, $\overline{n^3}/\overline{n}$
transparently decrease in amplitude, the period $P_{\rm{SF}}(^3P_2)$
grows very rapidly and $\varepsilon_\nu^{(\rm{B})}$ also decreases
conspiracy. Thus, the amplitude of the glitch also apparently
decrease . From Eqs.(24, 33), we obtain
\begin{equation}
  \frac{\Delta\Omega}{\Omega}\propto\frac{I_{\rm{core}}}{I_{\rm{crust}}}\cdot\frac{\Omega_{\rm{core}}}{\Omega}[\frac{\overline{n^3}}{10^2\overline{n}}]
(\frac{P_{\rm{SF}}(^3P_2)}{1\rm{ms}})^{-1},
 \label{34}
\end{equation}

For the same young pulsar, comparing the Eq.(32), we may find an
important statistical formula concerning the glitch amplitude
$\Delta\Omega/\Omega$ and the time interval between successive
glitches that can be motored
\begin{equation}
  \frac{\Delta\Omega}{\Omega}\propto\frac{I_{\rm{core}}}{I_{\rm{crust}}}\cdot\frac{\Omega_{\rm{core}}}{\Omega}[\frac{\overline{n^3}}{10^2\overline{n}}]
(\frac{P_{\rm{SF}}(^3P_2)}{1\rm{ms}})^{-1}\times \Delta
t_{\rm{interval}},
 \label{35}
\end{equation}
This statistical relation (see Fig.2) is actually consistent with
the radio pulsar period of the young pulsar PSR J0537-6910 (LMC) in
Magellanic clouds observed after a long period of 10 years
monitoring \citep{b36}. This is the strong sensational support to
our theory.

\textbf{We get the Eq.(35) to explain the observational phenomena.
Dr. Wang (He is the second author of the paper \citep{b36}) claim to
me that the observational data of monitoring of pulsar PSR
J0573-6910 are the most complete for the pulsar glitches. our model
may explain this phenomena that the pulsar PSR J0573-6910 has
glitches with amplitude change roughly proportional to the time
separation between two successive. As to most pulsars with glitches,
the question whether they have the similar rule  is still open due
to the monitoring data of the glitches of most pulsars.}

\subsubsection[]{Slow glitch phenomena}

According to Eq.(31) the duration time scale for glitch, we know
that for older pulsars, $\overline{n^3}/\overline{n}\approx (1-10)$,
their superfluid vortex quantum numbers decrease very fast. As
$\overline{n^3}/\overline{n}$ decreases and
$P_{\rm{SF}}(^3P_2)\rightarrow 0.1$s, duration time scale $\Delta
t_{\rm{persist}}$ may be as long as ($10^4-10^6$)s (possibility
longer than some days). This may correspond to slow glitch
phenomena. Thus, slow glitch is a natural result by our theory. It
is also an important observational evidence in favor of our theory.

\section{Discussions and conclussions}

Based on the recent works on the magnetars \citep{b44,b45}and on
magnetic dipole radiation from the $^3P_2$ neutron superfluid
vortices ($^3P_2$MDRA) in neutron stars \citep{b41}. we propose a
new model of Glitch for young pulsars by oscillation between B and A
phase of $^3P_2$ neutron superfluid. The main ideas are given
following:

1) On the direct Urca (DUrca) process:

For the young neutron stars, the magnetic field may be ultra strong
($B\gg B_{cr} =4.4\times10^{13}$Gauss). The DUrca process of the
neutron star cooling may happen in the ultra strong magnetic field,
although it is prohibited in the neutron stars with weaker magnetic
field ($B\ll B_{cr}$). This result is derived in our paper, which is
due to the Fermi energy of the electrons increasing with the
magnetic field ($E_{\rm{F}}(e)\propto (B/B_{\rm{cr}})^{1/4}$)
\citep{b44}.

2) A-B Phase oscillation of the $^3P_2$ neutron superfluid:

When the $^3P_2$ neutron superfluid is in the B -phase in the ultra
magnetic field, the heating rate by ($^3P_2$MDRA ) is far more than
the cooling rate by the DUrca process. By the heating process, the
ordered magnetic moments with the direction of the magnetic field
are gradually transformed into fully chaotic ESP state (A phase,
i.e., equal probability phase). Once the B phase of the $^3P_2$
neutron superfluid is translated into the A phase, the $^3P_2$MDRA
heating mechanism becomes more weaker than the direct Urca cooling
process. The cooling rate will make the $^3P_2$ neutron superfluid
again soon deviate from the ESP state, the A phase is translated
into to the B phase again. $^3P_2$ neutron superfluid again induced
magnetic moment, resulting in the corresponding induce magnetic
field. This is just A-B Phase oscillation.

3) Glitch mechanism:

During the heating process, some $^3P_2$ neutron Cooper pairs (the
fraction is $\zeta$, and $\zeta\ll 1$) are also disintegrated by
heating process and they will be broken up into normal neutrons at
the same time. That is, when the $^3P_2$ neutron superfluid is
translating towards The A phase, the fraction $\zeta$ of $^3P_2$
Cooper pairs are split into the normal neutrons simultaneously. The
total number of being disintegrated $^3P_2$ neutron Cooper pairs is
$\zeta N$ ($^3P_2$). For the total number of the normal neutrons,
the fraction of them is about $\zeta q (q\approx0.087)$.  When the
normal neutron component is accumulated to some amount (for an
example, the fraction reaches at $\zeta \sim 10^{-7}$), they are
strongly coupled with the normal protons by a nuclear force, which
are strongly coupled with the electrons in the crust and the shell
of the neutron star also by the Coulomb interaction. Due to these
strong couplings, the slowly rotating crust (and the shell) will be
suddenly driven by the internal rapidly rotating neutron superfluid.
That is, the suddenly strong couplings make the crust (and the
shell) suddenly rotating faster, or suddenly accelerating. This is
just the Glitch. The Glitches are a repeat phenomena with quasi
period $^3P_2$ neutron superfluid B phase $\Rightarrow$ A phase
$\Rightarrow$ B phase $\Rightarrow$ Many repeated Glitches with
quasi-period. With the repeating phase transition processes, the
vortex quantum number, $n$, of $^3P_2$NSV is gradually reduced, and
the heating rate $\varepsilon_\nu^{(\rm{B})}$ is also getting lower
and lower. After a number of glitch, the time intervals of
successive Glitch will gradually become long, and the amplitude of
the Glitch is downward. But there is no strict rule or following the
periodic or quasi periodic relation due to some random factor. The
heating rate $\varepsilon_\nu^{(\rm{B})}$ of the old neutron star
decrease lower and lower with decreasing of the vortex quantum
number, $n$, and with longing of the rotating period of the $^3P_2$
superfluid core. When the heating rate $\varepsilon_\nu^{(\rm{B})}$
of the old neutron star becomes lower than the cooling rate of the
DUrca, the $^3P_2$NSV  state is no longer returned to the normal
neutron fluid state. The phase oscillation of the system is stopped
immediately. That means that old pulsars will no longer present the
glitch.

4) Comparing with observation: the observation for the pulsars with
the period $P>0.7$s, and the glitch no has not been detected
\citep{b34,b35}. With the pulsar period increasing, the amplitude of
glitches decreases, and with the magnetic field weakening, the
amplitude of glitches decreases. The slowing glitch phenomenon for
some older pulsars is a naturally result in our theory. The
relationship between Glitch amplitude and the stationary time
interval (See the Fig.2, referenced from \citet{b36}) is naturally
got by our theory.

\acknowledgments

This work was supported in part by the National Natural Science
Foundation of China under grants 11565020, 10773005, and the
Counterpart Foundation of Sanya under grant 2016PT43, the Special
Foundation of Science and Technology Cooperation for Advanced
Academy and Regional of Sanya under grant 2016YD28, the Scientific
Research Starting Foundation for 515 Talented Project of Hainan
Tropical Ocean University under grant RHDRC201701, and the Natural
Science Foundation of Hainan Province of China under grant 118MS071.

\end{document}